# Identifying the Dirac point composition in $Bi_{1-x}Sb_x$ alloys using the temperature dependence of quantum oscillations

Joon Sang Kang[1,a)], Dung Vu[1] and Joseph P. Heremans[1,2,3]

1. Department of Mechanical and Aerospace Engineering

2. Department of Physics

3. Department of Materials Science and Engineering

The Ohio State University, Columbus, Ohio 43210

Dung Vu ORCID 0000-0001-9085-0436

Joseph P. Heremans ORCID 0000-0003-3996-2744

Author to whom correspondence should be addressed: kang.1284@osu.edu

**Abstract**

The thermal chiral anomaly is a new mechanism for thermal transport that occurs in Weyl semimetals (WSM). It is attributed to the generation and annihilation of energy at Weyl points of opposite chirality. The effect was observed in the $Bi_{1-x}Sb_x$ alloy system, at $x$=11% and 15%, which are topological insulators at zero field and driven into an ideal WSM phase by an external field. Given that the experimental uncertainty on $x$ is of the order of 1%, any systematic study of the effect over a wider range of $x$ requires precise knowledge of the transition composition $x_c$ at which the electronic bands at the L-point in these alloys have Dirac-like dispersions. At $x>x_c$, the L-point bands are inverted and become topologically non-trivial. In the presence of a magnetic field along the trigonal direction, these alloys become WSM's. This paper describes how the temperature dependence of the frequency of the Shubnikov-de Haas oscillations $F(x,T)$ at temperatures of the order of the cyclotron energy can be used to find $x_c$ and characterize the topology of the electronic Fermi surface. Alloys with topologically trivial bands have $dF(x,T)/dT>0$, those with Dirac/Weyl fermions display $dF(x,T)/dT<0$.

**Introduction**

Weyl semimetals (WSM) are characterized by an electronic band structure which has linearly dispersing bands that intersect at Weyl points (WPs) in a system that breaks time reversal symmetry (TRS) or inversion symmetry (IS). The WPs come in pairs located at specific points in $k$-space, $+\mathbf{k}_0$ and $-\mathbf{k}_0$ which are the source and sink of Berry curvatures.. The dispersion relation near the WPs is characteristic of massless particles and 3-dimensional:

$$E(\mathbf{k}) = \pm v_F \hbar \mathbf{k}, \ \mathbf{k} = (k_x, k_y, k_z) \tag{1}$$

Here, $v_F$ is the Fermi velocity, $\hbar$ is the Planck's constant, and $\mathbf{k}$ is measured relative to $\pm\mathbf{k}_0$ for the respective WP in the pair. The electrical and thermal transport properties of WSM display the chiral anomaly. First predicted by Nielsen and Ninomiya[1], the chiral anomaly manifests itself as an extra electrical conductivity that appears in the samples in the presence of parallel electric ($\mathbf{E}$) and magnetic ($\mathbf{B}$) fields applied in the direction of the WPs separation in $k$-space, i.e. the direction from $-\mathbf{k}_0$ to $+\mathbf{k}_0$. In the extreme quantum limit (EQL), when all electrons and holes are on the last Landau level, this additional electrical conductivity is given by:

$$\sigma_{zz}(B_z) = N_w \frac{e^3 v \tau}{4\pi \hbar^2} B_z \tag{2}$$

where $N_w$ is the number of degenerate pairs of WPs in the Brillouin zone, $e$ is the electron charge, $v$ is the electron velocity and $\tau$ is the inter-Weyl point scattering time. The thermal chiral anomaly[2] manifests itself as the creation of an additional electronic thermal conductivity that appears in the samples in the presence of parallel thermal gradient ($\nabla T$) and $\mathbf{B}$ applied in the direction of the WPs separation in $k$-space. In the extreme quantum limit, it is given by[2]:

$$\kappa_{zz}(B_z) = N_w T \frac{\pi^2}{3} \frac{k_B^2 e v \tau}{4\pi \hbar^2} B_z \qquad (3)$$

where $T$ is the temperature and $k_B$ is the Boltzmann constant. The additional electrical conduction manifests itself in WSM's as a negative longitudinal magnetoresistance (NLMR). Unfortunately, observing an NLMR is not a sufficient proof of the existence of the anomaly, because extrinsic effects caused by the applied magnetic field on the current distribution in the sample, most notably current jetting effects,[3,4] also give rise to an NMLR. The additional thermal conductivity is observed as a positive longitudinal thermal conductivity,[2] which is a more reliable indicator of the existence of the anomaly because no electrical current flows during a thermal conductivity measurements and because the lattice thermal conductivity evens out the heat distribution lines and is insensitive to field. In reference [2], the increase in both electrical and thermal conductivity were observed in high magnetic fields along applied along the trigonal ($z$) direction in $Bi_{1-x}Sb_x$ alloys with $x = 10.5 \pm 0.5$ at.%, and $15.1 \pm 0.7$ at.% (referred to nominally 11 and 15%).

The evolution of the band extrema in $Bi_{1-x}Sb_x$ alloys as a function of $x$ at zero field is shown schematically in Fig. 1[5]. In topologically trivial semiconductors or semimetals, the states with a symmetric wavefunction (here denoted s) constitute the conduction band, whereas the valence band consists of states with an antisymmetric (here denoted a) wavefunction. This is the case in $Bi_{1-x}Sb_x$ alloys for $x < x_c$. With increasing Sb content, the gap $E_g$ between the $L_s$ and $L_a$ bands closes, until it reaches zero at a concentration $x_c$ ($\approx 5 \pm 1$ at%[5]). The green line in Fig. 1 represents the minimum of a second valence band at the T-point of the Brillouin zone. Holes in that band have conventional parabolic dispersions and, at any value of $x$, are topologically trivial. For $x > x_c$ the bands are inverted: the conduction band has *a* symmetry (denoted $L_a$) and the valence band has *s* symmetry (denoted $L_s$). At $x > 8$ or 9%, the T-hole-band maximum falls below the $L_s$ valence band, and the alloys become direct-gap topological insulators, in fact, the first topological insulators identified experimentally by ARPES measurements.[6] At zero field, the 11% and 15% alloys are direct-gap semiconductors with gaps ($E_g \approx 13$ meV at $x = 11\%$ and $E_g \approx 30$ meV at $x = 15\%$) at the L-points of the Brillouin zone. In the presence of a strong magnetic field oriented along the trigonal z-axis, their band gap closes with increasing field[2] because of the extremely large Landé g-factor in these alloys.[7] Above a critical applied magnetic field $B_z$, the $x = 11\%$ and 15% alloys form ideal WSMs, meaning that the dispersions are given by Eq. (1) and the electrochemical potential $E_F$ is at the WP energy ($E_F = 0$) within the experimental energy range. The WPs are centered around the L-points of the Brillouin zone, and separated along a direction that is mostly aligned with the trigonal z-axis, so that $\kappa_{zz}(B_z)$ shows a strong thermal chiral anomaly, which gives rise to a very large increase in electronic thermal conductivity in magnetic field.[2] There also are no trivial bands at energy $E_F$ and no unintentional doping. Thus, $E_F$ is pinned to the WPs because those points have the minimum system density of states (DOS); an ideal WSM displays no Shubnikov-de Haas (SdH) oscillations. The Fermi surface of these ideal WSMs consists only of degenerate pairs of WPs with opposite Berry curvatures.

While the experiments in Reference [2] were carried out on ideal WSM's, only alloys with $9\% < x < 18\%$ are expected to fall in that category.[5,8] Above about $x \approx 18\%$, a new trivial valence band, the H-band, crosses the $L_s$ band, making the alloys indirect-gap semiconductors at zero field, and above $x \approx 22\%$ it crosses the $L_a$ band, making the alloys antimony-like semimetals.[5] In order to extend the thermal chiral anomaly to room temperature, it is necessary to investigate $Bi_{1-x}Sb_x$ alloys with as wide range of $x$ as possible because $Bi_{1-x}Sb_x$ is expected to give rise to a WSM phase in magnetic field as a function of $x$, even if those WSMs are not ideal. This means $E_F$ in them falls in a band and not at the WP's, and if their Fermi surfaces contain trivial pockets. Weyl physics is expected to remain even if $E_F$ falls inside, but within the bandwidth of, the Weyl bands. Here, we focus our work on alloys in the 0% < $x$ < 7% range.

Note that the experimental accuracy with which $x$ is reported in the literature is only about ±1%. Furthermore, because the parameters for band structure calculations are adjusted to reproduce the experimental data for the gaps[2], the calculated band structures also have about the same uncertainty. An experimental study is therefore necessary to determine which range of alloy compositions have topologically trivial and which have topologically non-trivial L-point bands, irrespective of the presence of a trivial T-point band. In this article, we describe a method for this condition based on the temperature-dependence of the frequency $F(T)$ of SdH oscillations. At $T=0$, $F(T=0)$ itself is a measure of the cross-sectional area of the Fermi surface $A(E_F) = (e/h) F$ normal to the direction of the applied field. At finite temperature, $F(T)$ probes the cross-sectional area of the Fermi surface $A(E_F(T), T)$ averaged over an energy range ~ $k_B T$ near $E_F$, as depicted by the circles on the bands in the insets in Fig 1. This potentially contains two mechanisms that give rise to $dF(T)/dT \neq 0$. First, there is a change in $E_F(T)$ with $T$, called the "Sommerfeld correction"[9], which gives $dF(T)/dT > 0$. The second mechanism is reported by Guo et al[10] and shown in that publication to be larger than the first in Dirac bands. At temperatures such that $2\pi^2 k_B T > \hbar \omega_C$, where $\omega_C$ is the electron cyclotron frequency, the temperature-dependence of the quantum oscillation frequency $F(T)$ is a measure of $-\left|\partial m_C / \partial E\right|$, the energy derivative of the cyclotron mass $m_c$. In topologically trivial bands with a parabolic dispersion the effective mass is constant which means $-\left|\partial m_C / \partial E\right|$ is zero. When the dispersion is Dirac-like and given by Eq. (1) (the case depicted in Fig 1 at $x = x_c$ range), the effective mass increases as $E$ moves away from the Dirac point, and the $-\left|\partial m_C / \partial E\right|$ correction term, which is proportional to $v_F^{-2}$, is much larger than the Sommerfeld correction.[10] What physically happens is that the heavier, higher-energy charge carriers are less likely to complete cyclotron orbits before encountering a phonon that perturbs the phase coherence of their wavefunction. The thermal average of the Fermi surface area $A(E, T)$ that gives rise to $F(T)$ is thus skewed toward lower-energy electrons in the Fermi distribution; these have a smaller Fermi surface cross-section so that the measured quantum oscillation $F(T)$ at finite temperature decreases with increasing $T$. The experimental test for an alloy composition to have reached $x = x_c$ is therefore to observe that $dF(T)/dT < 0$.

**Experiment**

A series of $Bi_{1-x}Sb_x$ samples with nominal concentration of antimony, ($x$ = 2.1, 3.3, 4.1, 5.3 and 7.2%) were prepared by the traveling molten zone (TMZ) method described elsewhere.[2] The sample properties are summarized in Table 1. The trigonal plane was identified visually and verified by X-ray diffraction (XRD). The nominal antimony concentration was obtained from these XRD spectra at 300 K and comparing the positions of the (009) peaks with the values given by Cucka and Barrett[11,] for alloys in the same composition range. The error bar was obtained by repeating the experiment multiple times and taking the standard deviation around the average value. The main source of error comes from small misalignments of the sample surface vis-à-vis the diffracting surface. The concentration was verified using X-ray fluorescence (XRF) based on the composition of polycrystalline alloys prepared by quenching. Table 1 shows that the two methods gave consistent results.

The samples were cut into cuboid shape of approximate dimensions 0.5×0.5×3 mm for transport measurements, with the sample long dimension along the trigonal (z) axis. Resistivity and Hall effect measurements were made using an AC bridge, Lake Shore 370, in a Quantum Design PPMS system using the AC resistivity/Hall puck, at temperatures from 300 to 2K and in magnetic fields of up to 7 Tesla.

**Results and discussion**

The convention used here for galvanomagnetic transport measurements of a resistivity labeled $\rho_{ij}(B_k)$ is that the first index (*i*) is that of the crystallographic direction of the current applied to the sample, the second (*j*) is the measured electric field, and the third (*k*) is the direction of the applied magnetic field. Thus, $\rho_{zz}(B_z)$ is the longitudinal magnetoresistance along the trigonal axis. $\rho_{zx}(B_y)$ the transverse Hall effect in bisectrix (*y*) magnetic field. The temperature-dependence of the zero-field trigonal resistivity is shown in Fig. 2a. The Hall resistivity $\rho_{zx}(B_y)$ at low field (-0.5T to 0.5T) is shown in Fig. 2b.

The magnetic-field dependence of the Hall resistivity $\rho_{zx}(B_y)$ is extremely non-linear, due to the simultaneous presence of electrons and holes,[12] This electron/hole compensation makes it impossible to derive the properties of the majority carrier from Hall measurements when the field is along the trigonal direction. However, in the low-field limit of $\rho_{zx}(B_y)$ it is possible to derive the carrier concentration and mobility of the carrier with the highest mobility (here the carriers in the L-point bands) from the equations in Ref. [13]:

$$n = \lim_{B_y \to 0} \frac{eB_y}{\rho_{zx}(B_y)}$$
$$\mu = \lim_{B_y \to 0} \frac{\rho_{zx}(B_y)}{\rho_{zz}(0)} \tag{4}$$

These values are shown as a function of temperature in Fig. 3. The mobility of all samples reaches several million cm$^2$/V-s at 10K, indicative of the excellent sample quality. The electron concentration in the 5.3% sample and the hole concentration in the 7.2% sample decrease steadily with decreasing $T$ and reach very low values at 10 K (shown in Table 1). Since there are about $6\times10^{22}$ atoms/cm$^3$ in Bi, a residual carrier concentration of $3.2\times10^{15}$ cm$^{-3}$ in the semiconducting 7.2% alloy indicates that the concentration of residual dopants in these alloys is of order of $5\times10^{-8}$ atom fraction. The most common residual impurities in Bi being Pb and Sn, both acceptors,[13] the p-type nature of the 7.2% alloy can be attributed to impurities in the starting materials. The mobility of the 7.2% sample below 30 K decreases with decreasing temperature, indicating ionized impurity scattering, which is also consistent with this hypothesis. The 7.2% alloy is thus a semiconductor and, at a temperature above 20 K where the charge carrier concentration becomes activated, is an intrinsic semiconductor. In the presence of a magnetic field, it is expected to become an ideal WSM, as did the samples in Ref [2].

The other alloys are n-type. The 2.1, 3.3 and 4.1% alloys' electron concentrations at 10K are in the high-$10^{16}$ cm$^{-3}$ range and temperature independent up to 20-30 K, which is an indication that they are semimetals in which the presence of the T-point band determines the position of $E_F$. This conclusion is again consistent with the observation that their mobility shows no trace of decreasing with decreasing temperature, indicating that if ionized impurities are present, their scattering effect is screened by free electrons that do not arise from doping. The charge neutrality condition in semimetals imposes that there are as many electrons as holes, to within the density of acceptor impurities. Because all samples were prepared from the same starting materials and using the same synthesis procedure, we can assume that this density is of order of $3\times10^{15}$ cm$^{-3}$ as in the semiconducting 7.2% alloy. In the semimetals, the low-field Hall effect still mostly measures the concentrations of electrons, reported in Table 1, because the electron mobility in the L-point conduction bands is much higher than that in the T-point hole band in Bi,[14] a situation that is reasonable to assume extends to the semimetallic $Bi_{1-x}Sb_x$ alloys that have similar band structures. The 5.3% sample represents an intermediate case.

The resistance and longitudinal magnetoresistance $R_{zz}(B_z)$ of the samples is reported in Fig. 4 at 2, 3, 5, 10, 15 and 20 K. The 2.1, 3.3 and 4.1% samples, which the Hall effect measurements reveal to be semimetals with $E_F$ in the L-point valence band, clearly show quantum oscillations at one single frequency, the SdH effect. The background shows a NLMR, but this cannot be taken as being a real physical magnetoresistance because Bi and its alloys are extremely prone to displaying current jetting[4] and only extraordinary precautions[2] can avoid this. The behavior of the 5.3% sample is ambiguous, as either the period of oscillations, if they exist, cannot be resolved in the fields available, or the sample reached the extreme quantum limit (EQL) already at 2 T, as do the semiconducting samples in Ref [2], or the observed features in the MR are due to current jetting. The 7.2% sample shows no oscillations at all.

After background subtraction, it is possible to Fourier transform the oscillations' frequency in $1/B$ and derive a frequency $F(x,T)$ at each concentration and temperature. The values for $F(x,T=2K)$ are reported in Table 1. The corresponding values for the Fermi surface cross-section $A(E_F,T)$ is also reported. In order to establish that the observed oscillations arise from the L-point electrons, and not the T-point holes, the following procedure was developed. In Bi, the T-point hole bands are parabolic and the Fermi surfaces are ellipsoids of revolution with an effective mass along the trigonal axis $m_{T,z} = 0.67\ m_e$ and masses in the trigonal plane of $m_{T,x} = m_{T,y} = 0.064\ m_e$ ($m_e$ is the free electron mass).[5] Assuming that the T-hole band masses remain the same in $Bi_{1-x}Sb_x$ for $x<5\%$, it is thus possible to calculate the hole concentration from the Fermi surface cross-section.

We obtain, for the 2.1% sample $p = 1.6\times10^{17}$cm$^{-3}$; for the 3.3 % sample $p = 1.2\times10^{17}$cm$^{-3}$, and for the 4.1 % sample $p = 7.1\times10^{16}$ cm$^{-3}$ numbers that are clearly not compatible with the densities obtained by the Hall measurements in Table 1. The lack of correspondence invalidates the hypothesis that the oscillations are due to T-point holes, and thus points to the oscillations being due to electrons at the L point. The L-point bands change strongly from more parabolic to Dirac-like with increasing $x$, so that assuming their constancy with $x$ is not an acceptable hypothesis.

It is now possible to plot the change of $F(x,T)$ as a function of temperature, which is done in Fig. 5. The theoretical prediction[10] that $dF(T)/dT < 0$ in an alloy of composition such that the L-point band has a Dirac dispersion, and not in trivial bands, is strikingly confirmed by the experiment. The error bars were determined from the full-width half maximum of the Lorentzian fit made through the Fourier transform of the data. It is also interesting to note that, while the alloys with trivial bands see a decay of this bandwidth, so much that the error bars on the 2.1 and 3.3% samples at 15 and 20 K are prohibitively large, the oscillations on the 4.1% sample remain well resolved even at 20 K. We submit that this experiment is another experimental proof of the validity of the theory in Ref.[10], and may even be somewhat more systematic than the proof offered there, because the authors in [10] had to compare data on completely different systems ($Cd_3As_2$, $LaRhIn_5$ and $Bi_2O_2Se$) whereas here we show the evolution of $dF(T)/dT$ within one system.

**Conclusions**

We experimentally show how the temperature dependence of the frequency of quantum oscillations can be used as a diagnostic tool for the Dirac nature of bands. The uncertainty in composition $x$ of $Bi_{1-x}Sb_x$ alloys is of the order of 1%; this makes it difficult to establish from the literature exactly at which composition $x_c$ the bands are Dirac-like, even if the literature reports a value around 5±1 at%. Therefore, in a systematic study of the thermal chiral anomaly as a function of $x$, it is necessary to establish $x_c$ experimentally. We prove here that the method of checking for $dF(T)/dT < 0$ is practical and functional. Future work includes correlating the observation of $dF(T)/dT < 0$ with that of the thermal chiral anomaly.

**Acknowledgements**

This work was supported in part (DV, JPH) by the NSF MRSEC at OSU, entitled the "Center for Emerging Materials", grant number DMR-2011876, and in part (JK, JPH) by the MURI entitled "Extraordinary electronic switching of thermal transport", grant number UTA21-000333

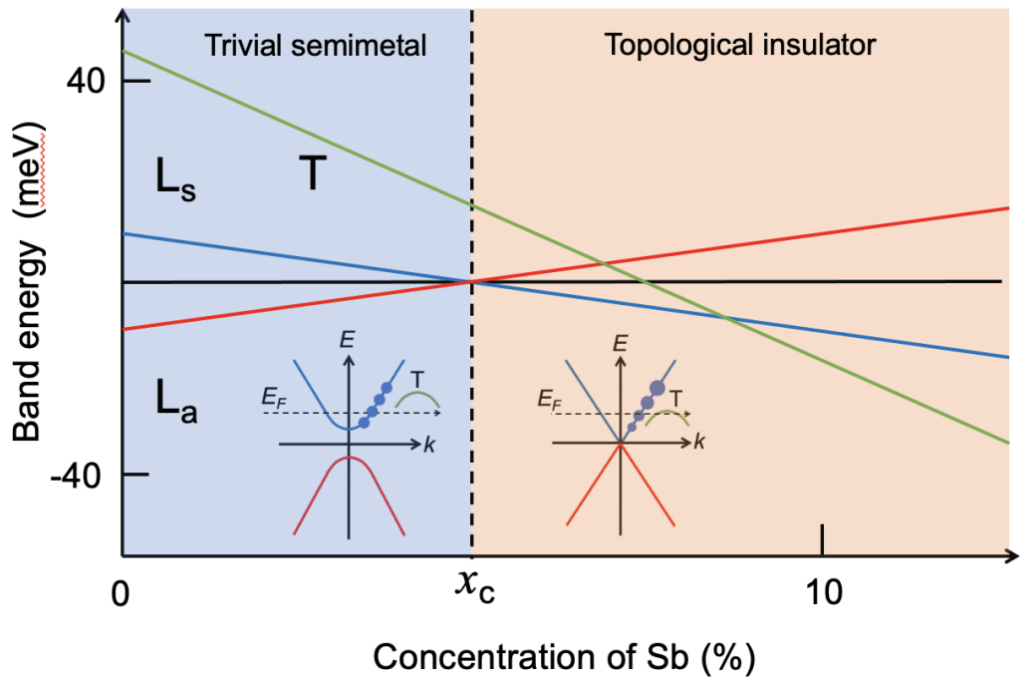

Fig. 1. Evolution of the band structure of $Bi_{1-x}Sb_x$ alloys as a function of antimony concentration $x$ (in at. %). The inserts show the dispersion relation in the topologically trivial phase ($x < x_c$) and at the composition $x_c$ at which the L-point bands have a Dirac dispersion. $x_c \approx 5\pm1$ at.%.[5] The blue dots schematically show the magnitude of effective mass of carriers distributed over $k_BT$ around $E_F$: it doesn't change in a parabolic band, but decreases with energy in a Dirac band. This gives rise to a change in the frequency of the quantum oscillations with temperature.

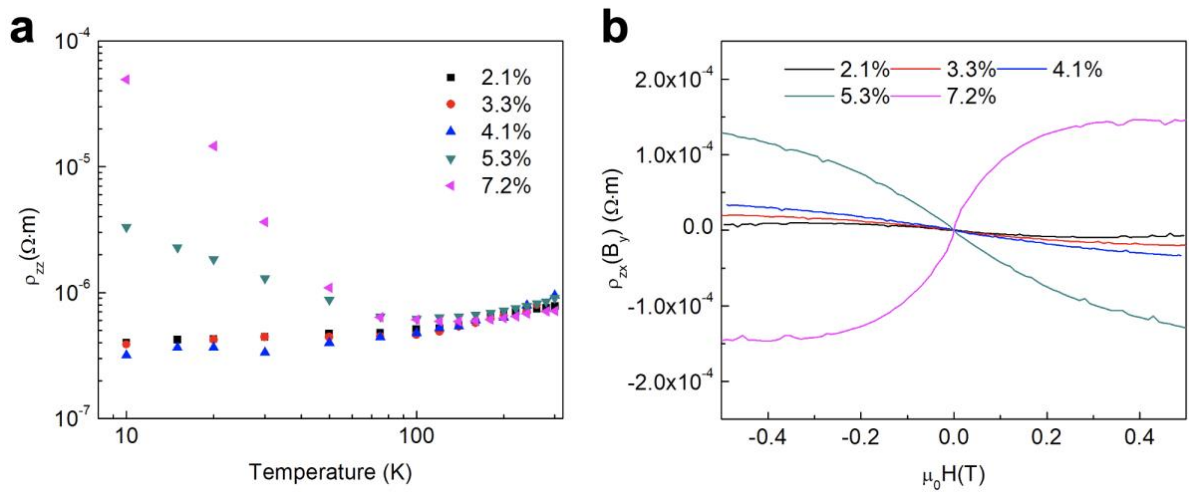

Fig. 2. Temperature dependence of: (a) the resistivity at zero magnetic field. (b) the Hall resistivity at 10K versus magnetic field, (b).

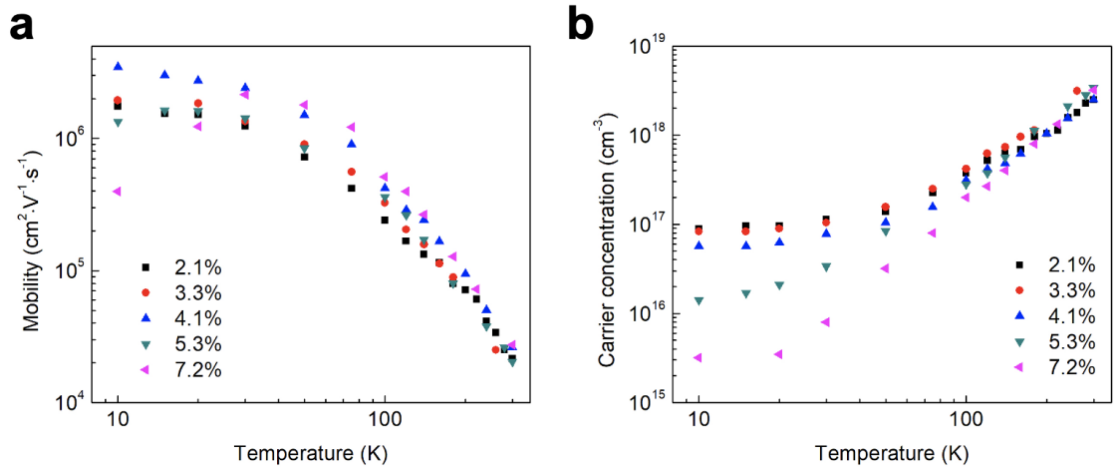

Fig. 3. Temperature dependence of: (a) the mobility along the trigonal direction versus temperature. (b) the low-field electron (samples 2.1-5.3%) or hole (7.2% sample) concentration.

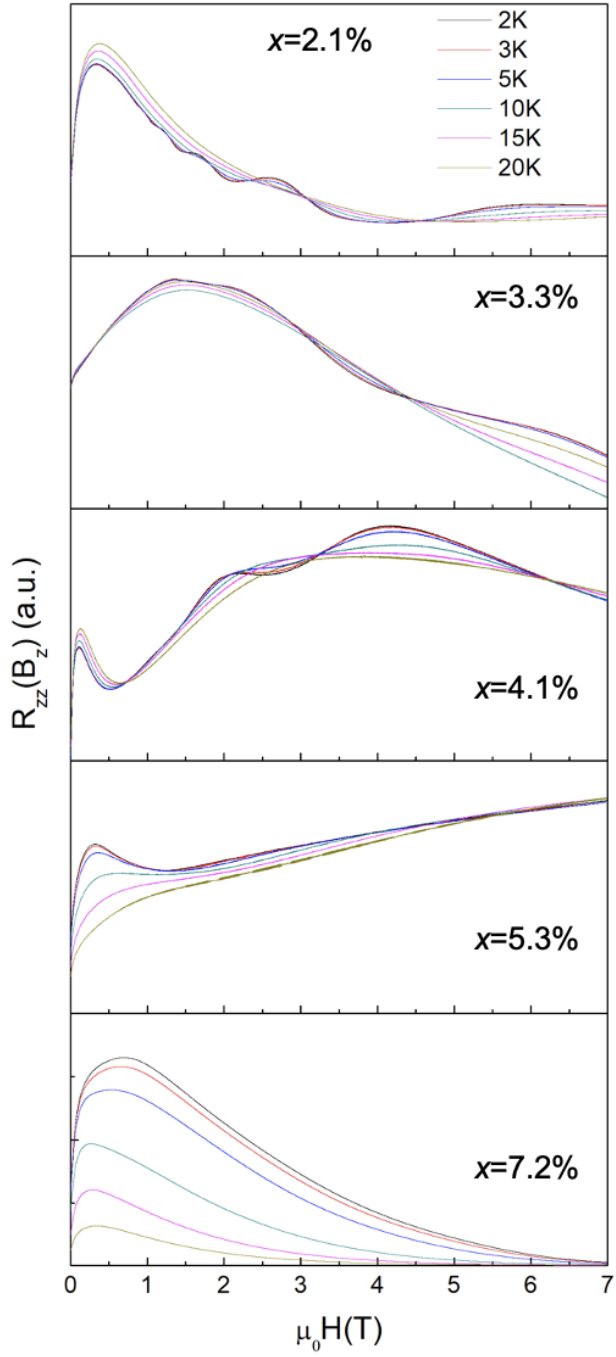

Fig. 4. Raw traces of $R_{zz}(B_z)$ in five different concentrations of $x$ showing the Shubnikov-de Haas oscillations. The absolute values of the resistivity are not reliable due to the current-jetting effects contributing to the observed decrease in resistivity with field.[3]

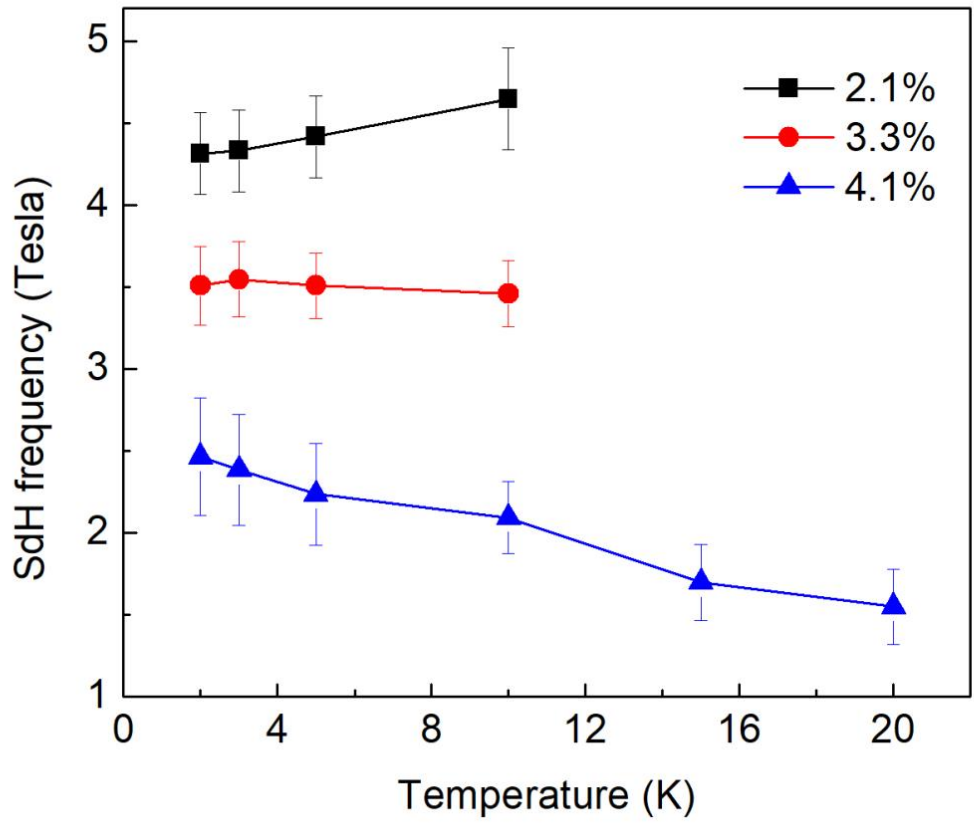

Fig. 5. Temperature-dependence of the Shubnikov-de Haas frequency for three different concentrations of *x*.

| Nominal $x$ (XRD, %) | 2.1±0.7 | 3.3±0.7 | 4.1±0.6 | 5.3±0.8 | 7.2±0.6 |
|---|---|---|---|---|---|
| Concentration $x$ (XRF, %) | 1.9±0.3 | 3.1±0.3 | 4.6±0.4 | 5.7±0.4 | 6.8±0.3 |
| Carrier concentration (cm$^{-3}$) at 10K | 8.9×10$^{16}$ (n) | 8.3×10$^{16}$ (n) | 4.5×10$^{16}$ (n) | 1.4×10$^{16}$ (n) | 3.2×10$^{15}$ (p) |
| Mobility (cm$^2$·V$^{-1}$·s$^{-1}$) at 10K | 1.8×10$^6$ | 1.9×10$^6$ | 3.5×10$^6$ | 1.3×10$^6$ | 4.9×10$^5$ |
| SdH frequency $F$ (Tesla) at 2K | 4.31±0.35 | 3.50±0.34 | 2.47±0.59 | - | - |
| Fermi surface area (m$^{-2}$) normal to the $z$-axis | 4.11±0.48 ×10$^{16}$ | 3.34±0.45 ×10$^{16}$ | 2.36±0.70 ×10$^{16}$ | - | - |

Table 1. Properties of the Bi$_{1-x}$Sb$_x$ samples studied here. The nominal concentration was obtained from X-ray diffraction, from the position of the (009) peaks. The concentration was double-checked by X-ray fluorescence. The carrier concentration and mobility was obtained from low-field measurements of the Hall coefficient $\rho_{zx}(B_y)$ and resistivity $\rho_{zz}$. The Shubnikov-de Haas frequency ($F$) is obtained in $R_{zz}(B_z)$.